\documentclass[aps,prd,preprintnumbers,nofootinbib]{revtex4}
\usepackage[dvipdfmx]{graphicx}
\usepackage{bm,latexsym,amsmath,amsthm,amssymb,amsfonts,color}

\usepackage[normalem]{ulem}

\begin{document}

\title{On uniqueness of static spacetimes with non-trivial conformal scalar field}


\author{${}^{1,2}$Yoshimune Tomikawa, ${}^{3,4}$Tetsuya Shiromizu and ${}^{4,3}$Keisuke Izumi}
\affiliation{${}^1$School of Informatics and Sciences, Nagoya University, Nagoya 464-8601, Japan}
\affiliation{${}^2$Faculty of Economics, Matsuyama University, Matsuyama 790-8578, Japan}
\affiliation{${}^3$Department of Mathematics, Nagoya University, Nagoya 464-8602, Japan}
\affiliation{${}^4$Kobayashi-Maskawa Institute, Nagoya University, Nagoya 464-8602, Japan}

\begin{abstract}%
We discuss the uniqueness of the static spacetimes with non-trivial conformal scalar field. 
Then, we can show that the spacetime is unique to be the Bocharova-Bronnikov-Melnikov-Bekenstein 
solution outside the surface composed of the unstable circular orbit of photon(photon surface). 
In addition, we see that multi-photon surfaces having the same scalar field values do not exist. 
\end{abstract}

\maketitle

\section{Introduction} \label{sec-intro}

The uniqueness theorem for static/stationary vacuum black holes is one of honorable consequences 
in general relativity \cite{Review}. This theorem guarantees comprehensive and definite predictions 
in black hole astrophysics/astronomy. 

Black holes in string theory are also interesting objects. One may want to consider the Einstein-scalar 
system as a simple model because scalar fields often appear in string theory. It is well known that there 
is a no-hair theorem by Bekenstein \cite{nohair}, that is, asymptotically flat and static black holes do not 
have regular scalar hair with non-negative potential. Interestingly, the 
Bocharova-Bronnikov-Melnikov-Bekenstein(BBMB) black hole solution \cite{bbm, bekenstein1974} exists 
in the Einstein-conformal coupled scalar field system. The metric of 
the spacetime is the same with the extreme Reissner-Nordstr\"{o}m black hole. This solution 
avoids Bekenstein's no-hair theorem because the scalar field is not regular at the event horizon. 
It is natural to ask whether the uniqueness of this black hole solution holds. In our previous work \cite{tomikawa2016}, 
we addressed this issue and then what we could prove is the uniqueness of the photon surface of 
the BBMB solution, not black hole. 
That is to say, the static spacetimes have to be the BBMB solution  
outside photon surface. The photon surface is defined as a generalization to the 
surface composed of the unstable circular orbit of photon in spherically symmetric black hole 
spacetimes (See Ref. \cite{Claudel} for the precise definition of the photon surface). A nontrivial 
point here is that we do not need the assumption of the existence of photon 
surface {\it a priori}. The requirement of the spacetime regularity gives the inner boundary of 
the region where we can discuss the uniqueness in the Einstein-conformal scalar field system. 
Eventually, the inner boundary turns out to be the photon surface \footnote{In classical level, the Einstein-conformal scalar field system(say, Jordan frame) 
is equivalent with the Einstein-massless scalar field system(say, Einstein frame) via a conformal 
transformation. In Ref. \cite{Y}, the uniqueness of the photon sphere (See the next paragraph) in the Einstein-massless 
scalar field system has been discussed. There, the existence of a photon sphere 
together with the constancy of the scalar field was assumed. In contrast, we will not assume the 
existence of such photon spheres {\it a priori}. Moreover, the spacetime in the Einstein frame has 
the curvature singularity at the locus of the photon sphere in the Jordan frame.}. 

The uniqueness of photon surface itself is also interesting. This is because the photon surface is outside of
black hole and then it can be observed. Therefore, if the uniqueness theorem holds for the region outside photon surface, it would 
be nice. The study for this direction was initiated by Cederbaum \cite{Cederbaum:2014gva} (See also Ref. 
\cite{Rogatko:2016mho} for other cases), but the argument is restricted to the photon surface with 
the constant time lapse function (called photon sphere). Although a perturbative study tells us 
positive result for the uniqueness \cite{yoshino2017}, it is hard to prove that without the additional 
requirement for the lapse function in non-perturbative level. Thus, our previous work for the 
Einstein-conformal scalar field system \cite{tomikawa2016} may give a hint 
to removing that assumption in general cases. In Ref. \cite{tomikawa2016}, we employed the Israel-type proof \cite{Israel}, 
that is the first version of the proof of the uniqueness theorem for static and vacuum black hole. Since the singleness of horizon 
is essential in the Israel-type proof, our previous proof is also restricted 
to cases corresponding to a single object. That is, we cannot remove the possibility of multi-photon 
surfaces system. Here we remind that, in a completely different way without the assumption of 
the singleness of horizon, the uniqueness of black hole has also been proven by Bunting and Masood-ul-Alam \cite{bm}. 
This means non-existence of multi-black holes in static vacuum spacetimes. Adopting the Bunting and 
Masood-ul-Alam type proof, the non-existence of multi-photon spheres has been proven for vacuum 
spacetimes \cite{Cederbaum:2015aha}.  

In this paper, employing Bunting and Masood-ul-Alam's way \cite{bm}, we discuss the uniqueness of the 
BBMB photon surface without the assumption of a single component of photon surface. 
With the scalar charge, there exist two kinds of photon surfaces characterized by the value of scalar field on each surface. 
We prove that, if only one of two kinds exists, the solution outside the surface is isometric to BBMB solution outside 
the photon surface. Thus, static multi-photon surfaces with the same scalar field values do not exist in the Einstein-conformal 
scalar field system. 

The rest of this paper is organized as follows. In Sec. \ref{setup}, we will describe the set-up and the BBMB solution. 
In Sec. \ref{SandL}, we will review a part of our previous work \cite{tomikawa2016} that we use in this paper; the relation between the scalar field and time lapse function, and the regularity conditions at the inner boundaries. 
In Sec. \ref{uniquenessofBBMB}, we will complete our proof. Finally, we will give the summary in Sec. \ref{summary}.

\section{Set-up and BBMB solution} \label{setup}

The theory that we consider is the Einstein-conformal scalar field system, whose action is 
written in
\begin{eqnarray}
S=\dfrac{1}{2\kappa} \displaystyle \int d^4 x\sqrt{-g} R - \displaystyle \int d^4 x
\sqrt{-g} \Big( \dfrac{1}{2} (\nabla \phi)^2 +\dfrac{1}{12} R\phi ^2 \Big). \label{BBMBaction}
\end{eqnarray}
Here, $\kappa:=8\pi G$, $\phi$ is the conformal scalar field and $R$ is the Ricci scalar. 

In this paper, we consider static spacetimes. The metric of static spacetime is written as 
\begin{eqnarray}
ds^2=-V^2(x^i)dt^2 +g_{ij}(x^k) dx^i dx^j,
\end{eqnarray}
where the Latin indices stand for the spatial components. 
We consider asymptotically flat spacetimes. As seen from linear perturbations around the Minkowski spacetime, 
we see that the metric near the spatial infinity behaves as 
\begin{eqnarray}
V=1-m/r+O(1/r^2),~~
g_{ij}=(1+2m/r)\delta_{ij}+O(1/r^2), 
\end{eqnarray}
where $r:=|\delta_{ij} x^i x^j|^{1/2}$ and $m$ is the Arnowitt-Deser-Misner(ADM) mass. 
$\lbrace x^i \rbrace $ are the asymptotically Cartesian coordinate near the spatial infinity. 
For the scalar field, we see the following asymptotic behavior;
\begin{eqnarray}
\phi=O(1/r). 
\end{eqnarray}
The above asymptotic behaviors for the metric and scalar field give us the boundary condition at the 
spatial infinity. 

The equation of motion for the scalar field is written in 
\begin{eqnarray}
D_i(VD^i \phi)=0,\label{scalar}
\end{eqnarray}
where $D_i$ is the covariant derivative with respect to 
the metric $g_{ij}$ of $\Sigma$.
The Einstein equation becomes
\begin{eqnarray}
\Bigl(1-\frac{\kappa}{6}\phi^2 \Bigr)D^2V=\frac{\kappa}{6} \Bigl[V(D\phi)^2+2\phi D^iV D_i \phi \Bigr] \label{E00}
\end{eqnarray}
and
\begin{eqnarray}
\Bigl(1-\frac{\kappa}{6}\phi^2 \Bigr)({}^{(3)}R_{ij}-V^{-1}D_i D_j V)
=\frac{\kappa}{6} \Bigl[4D_i \phi D_j \phi -g_{ij}(D\phi)^2-2\phi D_i D_j \phi \Bigr], \label{Eij}
\end{eqnarray}
where 
${}^{(3)}R_{ij}$ is the Ricci tensor with respect to the metric $g_{ij}$ of $t=$constant 
spacelike hypersurface. As see from the above equations, one has to impose the regularity 
condition at the surface $S_{p_\pm}$ specified by $\phi=\phi_{p_\pm}:=\pm {\sqrt {6/\kappa}}$ 
if one makes the spacetime regular. In general, $S_{p_\pm}$ can have the disconnected multi-components. 
Even for such cases, we simply write $S_{p_\pm}$.

It is known that there is a static and spherically symmetric black hole solution for the current theory 
\cite{bbm, bekenstein1974}. The metric and scalar field are given by 
\begin{eqnarray}
ds^2=-f(r) dt^2 + f^{-1}(r) dr^2 +r^2 d\Omega_2^2
\end{eqnarray}
and 
\begin{eqnarray}
\phi =\pm \sqrt{\dfrac{6}{\kappa}}\dfrac{m}{r-m},
\end{eqnarray}
where $f(r)=(1-m/r)^2$, $m$ is the mass of black hole and $d\Omega_2^2$ is the metric of the 
unit 2-sphere, that is, $d\Omega^2_2 =d\theta^2 +\sin ^2 \theta d\varphi^2$. The event horizon 
is located at $r=m$ and the scalar field diverges at there. This solution is called the BBMB 
solution. $S_{p_\pm}$ is located at $r=2m$ and coincides with the locus of 
the unstable circular orbit of photon.

We will prove that, if the spacetime has either $S_{p_+}$ or $S_{p_-}$($S_{p_+}$ or $S_{p_-}$ 
may have multi-components) and no horizon exists outside $S_{p_\pm}$, 
the asymptotically flat solution is unique as the BBMB solution. Then, it will be turned out that 
$S_p$($S_{p_+}$ or $S_{p_-}$) should have a single component. 

\section{Relation between scalar field and lapse function} \label{SandL}

In this section, we prove that, if a solution has either $S_{p_+}$ or $S_{p_-}$, 
there is a relation between the scalar field $\phi$ and lapse function $V$. 
Then, we will examine the regularity on $S_{p_+}$ (or $S_{p_-}$). The discussion basically follows 
our previous work~\cite{tomikawa2016}.

The assumption here is that the static solution has either $S_{p_+}$ or $S_{p_-}$. 
Without loss of generality, the solution has only $S_{p_+}$ because the theory is 
invariant under the flip of the sign of the scalar field $\phi$ ($\phi \leftrightarrow -\phi$). 
Hereinafter, $S_{p_+}$ is renamed $S_p$. 
We also assume that no horizon exists outside $S_{p}$ surfaces \footnote{
If the event horizon ($V=0$) encloses $S_p$, one can perform the conformal transformation 
so that the system becomes to the Einstein-massless scalar field system and then one can realize that 
non-trivial scalar fields cannot exist due to Bekenstein's no-hair theorem \cite{nohair}. 
Thus, the spacetimes should be the Schwarzschild spacetime \cite{Israel, bm}. 
There is still a room for the possibility that the event horizon and $S_p$ coexist, and that 
both $S_{p_+}$ and $S_{p_-}$ exist.
These cases are out of scope in this paper.
}.

From Eqs.~(\ref{scalar}) and (\ref{E00}), we can have an equation  
\begin{eqnarray}
D_i \big[ (1-\varphi) D^i \big((1+\varphi) V \big) \big]=0,
\label{eqa}
\end{eqnarray}
where $\varphi := \sqrt{\kappa/6} \phi$.

We consider the region $\Sigma$ satisfying $|\varphi|\le1$.  
Then, $\Sigma$ has two kinds of boundaries; the spatial infinity $S_{\infty}$ and the surfaces $S_p$($\varphi=1$). 
In general, $S_p$ is composed of multi-components. But, we simply write so. 
Integrating Eq. (\ref{eqa}) over $\Sigma$, we have
\begin{eqnarray}
0&=&\int_\Sigma D_i \big[ (1-\varphi) D^i \big((1+\varphi) V \big) \big] d \Sigma \nonumber \\
&=& \int_{S_{\infty}} (1-\varphi) D^i \big((1+\varphi) V \big) d S^i -\int_{S_p}(1-\varphi) D^i \big((1+\varphi) V \big) \nonumber \\
&=& \int_{S_{\infty}} D_i \big((1+\varphi) V \big) dS^i 
 \label{eqfirst}
\end{eqnarray}
In the above, we took the normal direction of $dS^i$ to be outward on $S_\infty$ and inward 
on $S_{p}$ to $\Sigma$. 

Next, we consider the following volume integration over $\Sigma$,
\begin{eqnarray}
0&=&\int_\Sigma (1+\varphi) V  D_i \big[ (1-\varphi) D^i \big((1+\varphi) V \big) \big] d \Sigma \nonumber \\
&=& - \int_\Sigma (1-\varphi) \big[ D \big((1+\varphi) V \big) \big]^2  d \Sigma +\int_{S_{\infty}} (1-\varphi^2) VD^i \big((1+\varphi) V \big)  dS^i-\int_{S_p} (1-\varphi^2) VD^i \big((1+\varphi) V \big)  dS^i\nonumber \\
&=& - \int_\Sigma (1-\varphi) \big[ D \big((1+\varphi) V \big) \big]^2  d \Sigma 
+\int_{S_{\infty}} D_i \big((1+\varphi) V \big) dS^i \nonumber \\
& = & - \int_\Sigma (1-\varphi) \big[ D \big((1+\varphi) V \big) \big]^2  d \Sigma \label{eqsecond}
\end{eqnarray}
In the above, we used the fact from the direct calculation that the third term in the second line 
vanishes. In the forth line, we used Eq. (\ref{eqfirst}). As stressed, we assumed 
that $S_p$ is composed of $S_{p_+}$ or $S_{p_-}$. If the both coexist, the third term in the second line 
does not vanish. Then, we see that $(1+\varphi) V$ is constant 
in $\Sigma$. The value of $(1+\varphi) V$ can be fixed by the asymptotic condition 
($(1+\varphi) V|_{r\to\infty}=1$), and then we have
\begin{eqnarray}
\varphi= V^{-1}-1. \label{pvrel}
\end{eqnarray} 
Note that $\phi=\phi_p$($\varphi=1$) corresponds to $V=1/2$. 


To examine the regularity at $S_p$, we look at the squares of the four dimensional Ricci 
and Riemann tensors 
\footnote{
In the current theory, the equations of motion give the vanishing of four dimensional Ricci scalar, $R=0$.
}. 
For the current case, we have \cite{tomikawa2016}
\begin{eqnarray}
R_{\mu\nu}R^{\mu\nu}  &=&  \frac{1}{\rho^4}+\frac{1}{(2V-1)^2\rho^2}\Bigl[ \Bigl(2(1-V)k_{ij}-\frac{1}{\rho}h_{ij} \Bigr)^2
\nonumber \\
& & +\Bigl(-2(1-V)k+\frac{1+2V}{\rho} \Bigr)^2+\frac{8(1-V)^2}{\rho^2}({\cal D}\rho)^2 \Bigr], \label{ricci2}\\
R_{\mu\nu\rho\sigma}R^{\mu\nu\rho\sigma} 
&=&  \frac{4}{V^2}D_i D_j V D^i D^j V+4{}^{(3)}R_{ij} {}^{(3)} R^{ij}-({}^{(3)}R)^2 \nonumber \\
&=&-\frac{4}{\rho^4}+\frac{4}{\rho^2}\Bigl[k_{ij} k^{ij}+\frac{2}{\rho^2}({\cal D}\rho)^2+\Bigl(k-\frac{1}{\rho} \Bigr)^2\Bigr]
\nonumber \\
& &~~~~+\frac{4}{(2V-1)^2\rho^2}\Bigl[ \Bigl(k_{ij}-\frac{1}{\rho}h_{ij} \Bigr)^2 +\frac{2}{\rho^2}({\cal D}\rho)^2+\Bigl(k-\frac{4V}{\rho} \Bigr)^2
\Bigr], \label{riemann2}
\end{eqnarray}
where $h_{ij}$ and ${\cal D}_i$ are the induced metric and the covariant derivative of $S_p$, respectively. 
Moreover, $k_{ij}$ is the extrinsic curvature of $S_p$ and $\rho:=|D \ln V|^{-1}$. In the above, we used the facts that, 
with the relation (\ref{pvrel}), Eq. (\ref{scalar}) becomes 
\begin{eqnarray}
D^2 V = \frac{1}{V}(DV)^2 \label{laplace}
\end{eqnarray}
and then the $(i,j)$-component of Einstein equation (\ref{Eij}) becomes
\begin{eqnarray}
\frac{2V-1}{V^2}\Bigl({}^{(3)}R_{ij}-\frac{1}{V}D_iD_jV \Bigr) =4\frac{D_iVD_jV}{V^4}-g_{ij}\frac{(DV)^2}{V^4}-2\Bigl(\frac{1}{V}-1 \Bigr)D_i D_j V^{-1}. \label{eij}
\end{eqnarray}
We would remind that the trace of Eq. (\ref{eij}) gives 
\begin{eqnarray}
{}^{(3)}R=\frac{2}{V^2}(DV)^2. \label{3riccidv2}
\end{eqnarray}
From Eqs. (\ref{ricci2}) and (\ref{riemann2}), we see that the spacetime is singular at $S_p$($V=1/2$) 
if the numerators do not vanish at $S_p$. Then, the regularity at $S_p$ implies \cite{tomikawa2016}
\begin{eqnarray}
k_{ij}|_{S_p}=\frac{1}{\rho_p}h_{ij} |_{S_p} \label{rc1}
\end{eqnarray}
and
\begin{eqnarray}
{\cal D}_i \rho|_{S_p}=0. \label{rc2}
\end{eqnarray}
The index ``$p$" indicates the estimation at $S_p$. At  each 
connected component of $S_p$, $\rho_p$ can have the different constant value.

\section{Uniqueness of BBMB spacetime} \label{uniquenessofBBMB}

Now, it is ready to discuss the uniqueness of the BBMB spacetime. We start from  
the two conformal transformations for $\Sigma$ as 
\begin{eqnarray}
\tilde g^\pm_{ij}=\Omega_\pm^2 g_{ij},
\end{eqnarray}
where we take the conformal factors as 
\begin{eqnarray}
\Omega_+=V
\end{eqnarray}
and
\begin{eqnarray}
\Omega_-=(1-V)^2/V.
\end{eqnarray}
Then, we have the two manifolds $(\tilde \Sigma^\pm, \tilde g^\pm)$. 

Using Eq. (\ref{3riccidv2}), it is  easy to see that the Ricci scalar for 
$\tilde g_{ij}^\pm$ vanishes as
\begin{eqnarray}
\Omega_\pm^2 {}^{(3)}\tilde R=2\Biggl[\Bigl(\frac{1}{V}
-\frac{\Omega'_\pm}{\Omega_\pm}\Bigr)^2-2\frac{\Omega''_\pm}{\Omega_\pm} \Biggr] (DV)^2=0, 
\end{eqnarray}
where the prime means the ordinary derivative with respect to $V$. 

Let $\tilde S^\pm_p$ be the inner boundaries of $\tilde \Sigma^\pm$. Then, 
we see that the metric $\tilde g^\pm$ and the extrinsic curvature $\tilde k^\pm_{ij}$ 
of $\tilde S_p^\pm$ become 
\begin{eqnarray}
\tilde g_{ij}^+|_{\tilde S_p^{+}}=\tilde g_{ij}^-|_{\tilde S_p^{-}}=\frac{1}{4}g_{ij}|_{S_p}
\end{eqnarray}
and 
\begin{eqnarray}
\tilde k_{ij}^\pm|_{\tilde S_p^{\pm}}=\pm \frac{1}{\rho_p}h_{ij} |_{S_p}, \label{2ffsp}
\end{eqnarray}
respectively. In the aboves, we used Eqs. (\ref{rc1}) and (\ref{rc2}). 
Thus, we can glue $(\tilde \Sigma^\pm, \tilde g^\pm)$ at $\tilde S_p^\pm$ without 
jump of the extrinsic curvature 
\footnote{There is a similar study on the photon sphere of the Schwarzschild spacetime \cite{Cederbaum:2015aha} 
(See also Refs. \cite{Rogatko:2016mho}). However, the two manifolds constructed via conformal transformation 
could not be glued at the photon sphere. To glue them continuously, the additional part made of the Schwarzschild 
spacetime was essential. 
}
and then we have $\tilde \Sigma=\tilde \Sigma^+ \cup \tilde \Sigma^-$ 
such that the Ricci scalar is zero even at $\tilde S_p^\pm$. 

Next, we look at the asymptotic behaviors on $(\tilde \Sigma^\pm, \tilde g^\pm)$;
\begin{eqnarray}
\tilde g_{ij}^+=\delta_{ij}+O(1/r^2)
\end{eqnarray}
and
\begin{eqnarray}
\tilde g_{ij}^-dx^idx^j  =  \frac{m^4}{r^4}\Big(1+O(1/r) \Bigr)
(dr^2+r^2d\Omega_2^2) 
=\Bigl(1+O(\bar r)\Bigr)(d\bar r^2+\bar r^2d\Omega_2^2),
\end{eqnarray}
where we set $\bar r:=m^2/r$. 
The spatial infinity in $\Sigma$ corresponds to a point $q$ in $\tilde \Sigma^-$, 
i.e., adding the point $q$, we can have $\tilde \Sigma \cup \lbrace q \rbrace$ with the 
zero Ricci scalar. 
Meanwhile, the ADM mass in $\tilde \Sigma \cup \lbrace q \rbrace$ vanishes. 
Since the Ricci scalar of $\tilde \Sigma \cup \lbrace q \rbrace$ is zero, 
the positive mass theorem \cite{pet} tells us that $\tilde \Sigma \cup \lbrace q \rbrace$ 
is flat. Thus, $(\Sigma,g)$ is conformally flat.

It is easy to see that $V^{-1}$ is the harmonic function in the flat space 
$(\tilde \Sigma^+,\delta)$ 
\begin{eqnarray}
\Delta_\delta V^{-1}=0, \label{lplc}
\end{eqnarray}
where $\Delta_\delta$ is the flat Laplacian. Since, as seen in Eq. (\ref{2ffsp}),
$\tilde S_p$ is totally umbilic surface 
in the flat space and $V^{-1}$ satisfying the flat Laplace equation (\ref{lplc}) is constant on $\tilde S_p$, 
we see that $\tilde S_p$ is spherically symmetric surface. 
Then, because of Eq. (\ref{lplc}) with the spherically symmetric boundary, all $V=$constant surfaces are spherically 
symmetric in $\tilde \Sigma^+$. Therefore, $(\Sigma, g)$ is also spherically symmetric because 
the conformal factor $\Omega$ depends only on $V$. It reminds us that there is no room for the possibility of 
multi-photon spheres.
It is known that the spherically symmetric solution is unique to be the BBMB solution \cite{xz}. 
Moreover, $S_p$ 
corresponds to the unstable circular orbit of photon at $r=2m$. Finally we can 
conclude that $\Sigma$ is isometric to the outside region of the photon sphere of the BBMB solution.

\section{Summary} \label{summary}

In this paper, we proved that the outside region of $S_p$ is unique as the BBMB solution 
in the Einstein gravity with the conformal scalar field. 
The proof can be applied the cases where no horizon exists outside $S_p$ and $S_p$ is composed only of $S_{p_+}$ or only of $S_{p_-}$. 
As a consequence, $S_p$ is the surface 
composed of the unstable circular orbit of photon (photon sphere) in the BBMB solution. We also 
saw that no static multi-photon surfaces systems exist under the above assumptions. 
We would stress that we did not assume the 
existence of the photon sphere and singleness of object 
{\it a priori}. 

We could not address the geometry of the inside region of the photon sphere. This remains to be future 
issue.

\begin{acknowledgments}
This work is initiated by the collaboration with Mr. K. Ueda. 
T. S. is supported by Grant-in-Aid for Scientific Research from Ministry of Education, 
Science, Sports and Culture of Japan (16K05344).
\end{acknowledgments}

\end{document}